\documentclass{article}
\usepackage{amsmath}

\begin{document}
\begin{center}
{\bf
A RECURSION  TECHNIQUE FOR DERIVING RENORMALIZED
PERTURBATION EXPANSIONS FOR ONE-DIMENSIONAL
ANHARMONIC OSCILLATOR  }\vspace{20pt}
{ \footnote{PACS Numbers 03.65G, 03.65S}}\\
I. V. DOBROVOLSKA  \\
and  \\
R. S. TUTIK{ \footnote{To whom correspondence should be addressed.
                    E-mail address: tutik@ff.dsu.dp.ua}}
\\
{ \small \it Department of Physics, Dniepropetrovsk State
University,} \\ { \small \it Dniepropetrovsk, 49050, Ukraine}

\vspace{20pt}

\begin{minipage}{0.9 \textwidth}
{ \footnotesize A new recursion procedure for deriving
renormalized perturbation expansions for the one-dimensional
anharmonic oscillator is offered. Based upon the
$\hbar$-expansions and suitable quantization conditions,  the
recursion formulae obtained have the same simple form both for
ground and excited states and can be easily  applied to any
renormalization scheme. As an example, the renormalized
expansions for the sextic anharmonic oscillator are considered. }
\end{minipage}
\end{center}

\section*{ \normalsize 1. Introduction}

In the past few decades intensive investigations have been carried
out on the one-dimensional anharmonic oscillator because of both
its role in the modeling of quantum field theory and its
usefulness in atomic, and molecular physics $^{1-4}$.  The
conventional way to study the energy eigenvalues and
eigenfunctions of this bound-state problem is the practical
application of perturbation expansions in a coupling constant. But
the nonconvergence of obtained expansions $^{5,6}$ compels us to
resort to modern procedures of summation of divergent series. Ones
of the most common among them are various versions of the
renormalization technique $^{7-9}$. The inclusion of one or more
free parameters corrects for the above mentioned deficiency, by
controlling and accelerating the convergence of the expansions not
only for eigenvalues but for eigenfunctions as well $^{10,11}$.

However, to provide a reasonable accuracy this method needs to
know high orders of the perturbation series. Unfortunately we can
easily obtain the high-order corrections only in the case of
ground states when the simple recursion relations of the
logarithmic perturbation theory $^{12-18}$ or the Bender-Wu
\mbox{method $^{5}$} for performing the Rayleigh-Schr{\"o}dinger
perturbation expansion are applied. The description of radially
excited states involves including the nodes of wave function  in
consideration that makes these approaches extremely cumbersome
and, practically, inapplicable.

Very recently, a new  semiclassical technique for deriving results
of logarithmic perturbation theory within the framework of the
one-dimensional Schr{\"o}dinger equation has been proposed
$^{19}$. Based upon an $\hbar$-expansion this technique leads to
recursion formulae having the same simple form both for the ground
and excited states.

The object of this paper is to extend the above mentioned
formalism and to adapt it to the treatment of any renormalization
scheme in terms of handy recursion relations within the framework
of the united approach. To this end the paper is presented as
follows. Section 2 contains a general discussion and the necessary
assumptions in the semiclassical treatment of logarithmic
perturbation theory. In Sect.3 quantization conditions obtained
are used for deriving the recursion formulae for perturbation
expansions. Sect.4 focuses on a derivation of the recursion
formulae for renormalized perturbation expansions. Sect.5 gives
the example of explicit application of proposed technique.
Finally, Sect.6 contains concluding remarks.

\section*{\normalsize 2. Semiclassical treatment of logarithmic perturbation
theory}

In this section, our goal is to describe a straightforward
semiclassical approach to obtaining results of the logarithmic perturbation theory.

We shall confine our interest to bound-state problem for the Schr{\"o}dinger equation
in one degree of freedom
\begin{equation}\label{1}
-{{\hbar}^2\over {2m}}U''(x)+V(x)U(x)= E U(x).
\end{equation}

We shell study the discrete piece of the spectrum related to the
potential function, $V(x)$, having a single simple minimum and can
be given by expression
\begin{equation}\label{2}
V(x)=\frac{1}{2}m\omega^{2}x^{2}+\sum_{i{\geq}1}f_{i}x^{i+2}.
\end{equation}

However, if the potential has a finite number of local minima,
the proposed below technique also can be applied. In this case
the obtained eigenvalues are the energies of quasidiscrete states
in a vicinity of each local minimum when the tunneling can be
neglected.

We want to restore the results of logarithmic perturbation theory
 for the $n$th eigenfunction and corresponding energy eigenvalue
by means of the explicit semiclassical expansions in powers of the
 Planck constant, $\hbar$.

Indeed, after the scale transformation, $x \to \sqrt{\hbar}x$, the
coupling constants, $f_i$, appear in common with powers of
Planck's constant. Therefore, the perturbation series must be
semiclassical $\hbar$-expansion, too.

It has been proved $^{20}$ that the energy eigenvalues under
 consideration should be concentrated near the minimum of the
 potential and should behave as
\begin{equation}\label{2.5}
E=\hbar\omega(n+1/2)+\sum_{k\geq 2}{E_k (\omega, n) \hbar^k}.
\end{equation}
A few procedures are elaborated for computing the coefficients
 $E_i(\omega, n)$. They involve, in particular, applying the
 methods of the comparison equation $^{21}$ and complex 'sprout' $^{22}$;
 an analytic continuation in the $\hbar$-plane $^{23}$;
various approaches within the framework of the WKB-approximation $^{24-28}$;
 quantization using the methods of classical mechanics $^{26,29}$;
 and, lastly, expansions in the $\hbar^{1/2}$-series $^{30}$.

Here we describe a new, simpler and more straightforward
 semiclassical technique $^{19}$. It involves two successive
 stages: first, we imply that the eigenvalue selection is
 effectively achieved by enforcing not the square integrability
 of the eigenfunction, but its analyticity, with applying the
quantization condition; and second, we state such a classical
 analog of the system, in the limit $\hbar \to 0$, which is
 appropriate for consideration of the low-lying levels.

To begin with, we use the substitution, $C(x)=\hbar U'(x)/U(x)$,
accepted in logarithmic perturbation theory, and rewrite Eq.
(\ref{1}) in the Riccati form
    \begin{equation}\label{3}
    \hbar C'(x)+C^{2}(x)=2m[V(x)-E].
    \end{equation}
It will be recalled that we attempt to solve this equation
 in a semiclassical manner with the following asymptotic
series expansions
\begin{equation}\label{4}
E=\sum^{\infty}_{k=0}{E_{k}\hbar^{k}},\;\;\;
C(x)= \sum^{\infty}_{k=0}C_{k}(x)\hbar^{k}.
\end{equation}
On substituting these expansions into the Riccati equation
 (\ref{3}) and collecting coefficients of the same powers
 of $\hbar$, one obtains the system
\begin{eqnarray}\label{9}
C_{0}^{2}(x) & = &2mV(x),\nonumber\\
C'_{0}(x)+2C_{0}(x)C_{1}(x) & = & -2mE_{1},\nonumber\\
\cdots \nonumber \\C'_{k-1}(x)+\sum_{i=0}^{k}C_{i}(x)C_{k-i}(x) & =
&-2mE_{k}.
\end{eqnarray}

For nodeless states this system coincides with those
 derived by means of the standard perturbation approach
 and can be solved straightforwardly. However, when
radial excitations are described within the framework
 of logarithmic perturbation theory the nodes of
 wave functions are included in some separate factor
and consideration becomes extremely cumbersome.
 We intend to circumvent this difficulty by making use
 of the quantization condition, inspired by the
 WKB-approximation $^{31,32}$.

Generally speaking, such the quantization condition
 simple means monitoring whether eigenfunction $U(x)$
 is analytic inside a closed contour. We would remind
 that a necessary condition for analyticity is
\begin{equation}\label{9.5}
\frac{1}{2\pi i}\oint{\frac{U'(x)}{U(x)}}dx = n,
\end{equation}
which must be a non-negative integer, giving the number of zeros
of $U(x)$ which lie inside the contour of integration. Hence, the
transition to the logarithmic derivative, $C(x)$, allows the nodes
of wave functions to be easily involved in consideration.

Taking into account that due to the Sturm-Liouville theory the
eigenfunction of the $n$th radially excited state has precisely
$n$ zeros on the real axis, we have
\begin{equation}\label{5}
\frac{1}{2\pi i}\oint{C(x)\,d x}= \frac{1}{2 \pi i }
\sum^{\infty}_{k=0}{\hbar^{k}\oint{C_{k}(x)\,d x}}=\hbar n, \;\;\;
n=0,1,2,\; ...
\end{equation}
where a contour of integration encloses the nodes of wave function
and no other singularities.

The above quantization condition must be supplemented with the
rule of achieving a classical limit for its right-hand side.
Unfortunately, under the influence of standard textbooks on
quantum mechanics, the semiclassical $\hbar$-expansions are
usually associated solely with the WKB-approach, for which the
rule of achieving a classical limit is
\begin{equation}\label{6}
\hbar\to 0,\;\;\;\; n\to\infty ,\;\;\;\;\hbar n={\rm const}.
\end{equation}
That implies that the WKB method is more suitable for obtaining
energy eigenvalues in the limiting case of large quantum numbers.

The semiclassical treatment of the low lying levels demands the
use of the different, as compared with the WKB-approach, rule of
achieving a classical limit, namely
\begin{equation}\label{7}
\hbar\to 0,\;\;\;\;n\sim O(1), \;\;\;\;\hbar n\to 0.
\end{equation}

This rule does provide lowering a particle to the bottom of a
potential well as $\hbar\to 0$ and was formerly applied in
deriving coefficients of the 1/N-expansions $^{33-36}$. As it
will be shown below, in one dimension case this way lead us to
restoring the logarithmic perturbation series.

Thus, in view of the new rule of passing to the classical limit
(\ref{7}), the quantization conditions (\ref{5}) are rewritten in
the final form as
\begin{equation}\label{8}
\frac{1}{2\pi i}\oint{C_{1}(x)\,d x}=n,\;\;\;\;\; \frac{1}{2\pi
i}\oint{C_{k}(x)\,d x}=0,\;\;\; k\not= 1.
\end{equation}

A further application of the theorem of residues to the explicit
form of functions  $C_{k}(x)$ easily solves the problem of taking
into account nodes of the wave functions for excited states.

\section*{\normalsize 3. Recursion formulae for  perturbation expansions}

We proceed now to deriving the recursion relations for obtaining
the $n$th eigenfunction and corresponding energy eigenvalue.

Let us consider the system (\ref{9}) and investigate the behaviour
of the functions $C_k(x)$. From the first equation it is seen that
$C_0(x)$ can be written in the form

\begin{eqnarray}\label{10}
C_{0}(x) = -{[2mV(x)]}^{1/2}
               = -m\omega x({1+{2\over{m\omega^2}}
                 \sum_{i\geq 1}{f_{i}x^i}})^{1/2}
               =  x\sum^{\infty}_{i=0}{C^{0}_{i}x^i} ,
\end{eqnarray}
where the minus sign is chosen from boundary conditions,  and
coefficients $C^{0}_{i}$  are defined by parameters of the
potential through the relations
\begin{eqnarray}\label{11}
C^{0}_{0} & = & -m\omega ,\;\;
\nonumber \\
C^{0}_{i} & = & {1\over{2m\omega}}
({\sum_{p=1}^{i-1}{C^{0}_{p} C^{0}_{i-p}-2mf_{i}}}) ,
\;i\geq 1.
\end{eqnarray}

From (\ref{2.5}) we recognize  that $C_0(0)=0$ and, consequently,
the function $C_{1}(x)$ has a simple pole at the origin, while
$C_{k}(x)$ has a pole of the order of  $(2k-1)$. Thus, $C_{k}(x)$
can be represented by the Laurent series
\begin{equation}\label{12}
C_{k}(x)= x^{1-2k}\sum^{\infty}_{i=0}{C^{k}_{i}x^i},\;\; k\geq 1.
\end{equation}
With definition of residues, this expansion  permits us  to
express the quantization conditions (\ref{8}) explicitly  in terms
of the coefficients $C^k_i$ as

\begin{equation}\label{13}
C^{k}_{2k-2}=n\delta_{1,k},
\end{equation}
where the  Kronecker delta is used. Thereby the common
consideration of the ground and excited states has indeed proved
to be possible.

The substitution of the series (\ref{11}) and  (\ref{12}) into
(\ref{9}) in the case $i \not= 2k-2 $ yields the recursion
relation for obtaining the coefficients $C^{k}_{i}$ :
\begin{equation}\label{14}
C^{k}_{i}=-{1\over{2C^{0}_{0}}}
[{(3-2k+i) C^{k-1}_{i }+\sum_{j=1}^{k-1}\sum_{p=0}^{i}
C^{j}_{p}C^{k-j}_{i-p}
+2\sum_{p=1}^{i}C^{0}_{p}C^{k}_{i-p}}] ,
\end{equation}
whereas if $i=2k-2$ we would find the recursion formula for the
energy eigenvalues
\begin{equation}\label{15}
2mE_{k}=- C^{k-1}_{2k-2 }-
\sum_{j=0}^{k}\sum_{p=0}^{2k-2} C^{j}_{p}C^{k-j}_{2k-2-p}.
\end{equation}

It is easy to verify that proposed technique does indeed restore
the results of the logarithmic perturbation theory. Upon
calculating first corrections to the energy eigenvalues we arrive
at the form  familiar from standard textbooks $^{24}$
\begin{equation}\label{15.5}
E=\hbar \omega(n+{\tfrac{1}{2}}) -
\hbar^2{\frac{15f^{2}_{1}}{4m^{3}\omega^{4}}}
(n^{2}+n+{\tfrac{11}{30}})+\hbar^2
{{3f_{2}}\over{2m^{2}\omega^{2}}}( n^{2}+n+{\tfrac12})+O(\hbar^3),
\end{equation}
with the obvious constraints
\begin{equation}\label{15.55}
n+{\tfrac{1}{2}}\ll \frac{2 m^2 \omega^3}{3\hbar f_2}, \frac{4 m^3
\omega^5 }{15\hbar f_1^2 }.
\end{equation}

The equations (\ref{14}) and
(\ref{15}) have the same simple  form both  for ground and excited
states and provide, in principle, the calculation of the
perturbation corrections up to an arbitrary order in the
analytical or numerical form.

Another salient feature of this technique is the fact that for
obtaining corrections to the energy eigenvalues and eigenfunctions
we need not have initially the explicit form of the exact solution
for the unperturbed potential. This solution is found
automatically in the process of solving recursion formulae.

Moreover, in this way the quasi-exact solutions may be obtained as
well. Indeed, to be more concrete, let us consider the ground
state for the anharmonic potential
\begin{equation}\label{100}
V(x)=V_2\; x^2+V_4\; x^4+V_6\; x^6
\end{equation}

Then the lowest order corrections derived by means of formulae
(\ref{14}) and (\ref{15}) are written as
\begin{eqnarray}\label{101}
E_1&=&{\sqrt{{V_2}}},\nonumber \\
E_2&=&{\frac{3\,{V_4}}{2\,{V_2}}},\nonumber \\
E_3&=&{\frac{3\,\left( -7\,{{{V_4}}^2} + 5\,{V_2}\,{V_6} \right) }{
      4\,{{{V_2}}^{{{5}/{2}}}}}},\nonumber \\
E_4&=&{\frac{-9\,\left( -37\,{{{V_4}}^3} + 40\,{V_2}\,{V_4}\,{V_6}
\right) }{
      8\,{{{V_2}}^4}}}, \\
E_5&=&{\frac{-15\,\left(
          2059\,{{{V_4}}^4} - 2992\,{V_2}\,{{{V_4}}^2}\,{V_6} +
            466\,{{{V_2}}^2}\,{{{V_6}}^2} \right) }{
      64\,{{{V_2}}^{{{11}/{2}}}}}},\nonumber \\
E_6&=&{\frac{9\,\left(
          101859\,{{{V_4}}^5} - 186380\,{V_2}\,{{{V_4}}^3}\,{V_6} +
            61420\,{{{V_2}}^2}\,{V_4}\,{{{V_6}}^2} \right) }{
      128\,{{{V_2}}^7}}},\nonumber
%
\end{eqnarray}
setting
\begin{equation}\label{102}
V_2=V_4^2/(4\; V_6)-3 \sqrt{V_6}
\end{equation}
immediately leads to the analytical expression for the ground
state energy and the wave function $^{37}$
\begin{equation}\label{103}
E=V_4/(2 \sqrt{V_6})
\end{equation}
and
\begin{equation}\label{104}
\Psi (x)=\exp(-V_4\; x^2/(4 \sqrt{V_6})- \sqrt{V_6}\; x^4/4)
\end{equation}
where for the simplicity we put $\hbar = 2 m = 1$.

\section*{\normalsize 4. Renormalized perturbation expansions}
The renormalization of perturbation expansions is based on a
redistribution of values  of series items for a partial summation
of them and  improving the convergence of expansions. From
mathematical point of view it means the transition from the
asymptotical expansion according to Poincare to the asymptotical
expansion according to Erd\'{e}lyi $^{38}$. Characteristic of this
method is that the zeroth approximation involves a set of
artificial parameters, which are not contained in the original
Hamiltonian and are determined order by order after calculating
the physical quantity perturbatively.

For the discussed one-dimensional anharmonic oscillator, as a
trial parameters we can choose a frequency and a mass. Though
proposed technique is easily applied to the mass renormalization
we do consider only the case of renormalization of a frequency as
being more physically motivated. For this purpose it is enough to
think of the harmonic oscillator frequency incoming in the
potential as a function of Plank's constant variable, with
subsequent its expansion in an $\hbar$-series $^{29}$.
However, for later use it is more convenient to take
\begin{eqnarray}{\label{16}}
\omega^2=\sum\limits_{k=0}^{\infty}\omega_k^2 \hbar^k .
\end{eqnarray}
The recursion system (\ref{9}) then reads
\begin{eqnarray}\label{17}
C_{0}^{2}(x) & = & m^2\omega^{2}_{\,0}x^{2}+2m \sum_{i\geq 1}f_{i}x^{i+2},
\nonumber\\
C'_{0}(x)+2C_{0}(x)C_{1}(x) & = & m^{2}\omega^{2}_{\,1}x^{2}-2mE_{1},\nonumber
\\ \cdots \nonumber\\
C'_{k-1}(x)+\sum_{i=0}^{k}C_{i}(x)C_{k-i}(x) & = &
m^{2}\omega^{2}_{\,k}x^{2}-2mE_{k},
\end{eqnarray}
with
\begin{equation} \label{18}
C^{0}_{0}=-m\omega_{0} ,\;
\;C^{0}_{i}={1\over{2m\omega_{0}}}
({\sum_{p=1}^{i-1}{C^{0}_{p} C^{0}_{i-p}-2mf_{i}}}),
\;i\geq 1.
\end{equation}
In the case $k \geq 1$, when $i \not= 2 k - 2$ and $i = 2 k - 2$ that
results in
\begin{eqnarray}{\label{19}}
C^{k}_{i} &=&
-{1\over{2C^{0}_{0}}}
\Bigg[ {- m^{2}\omega^{2}_{\,k}\delta_{i,2k}+(3-2k+i) C^{k-1}_{i } }
\nonumber \\ &&
+\sum_{j=1}^{k-1}\sum_{p=0}^{i} C^{j}_{p}C^{k-j}_{i-p}
+2\sum_{p=1}^{i}C^{0}_{p}C^{k}_{i-p}\Bigg] ,\nonumber
\\
2mE_{k}&=&- C^{k-1}_{2k-2 }-
\sum_{j=0}^{k}\sum_{p=0}^{2k-2} C^{j}_{p}C^{k-j}_{2k-2-p}\;, \end{eqnarray}
respectively. Here the coefficients $\omega_k^2$ are defined  by the chosen
version of renormalization.

In practice, the one-parameter schemes are usually used. They are
obtained
with the truncation of series (\ref{16}) as
\begin{eqnarray}{\label{20}}
\omega^2=\omega_0^2+\omega_k^2 \hbar^k ,
\end{eqnarray}
where  $\omega_0$ is a trial frequency and an order  in $\hbar$ of
the remainder  is determined by the anharmonicity of a potential
(\ref{2}).

A free parameter, $\omega_0$, is order-depended. At any finite
order of approximation, N, its optimal value is picked up by the
condition of  "minimal difference" $E_N (\omega_0) = 0$ $^{39,
40}$, and "minimal sensitivity", $(d/d\omega_0) \sum_{i=0}^{N} E_i
(\omega_0) = 0$  $^{10, 41}$ , because perturbatively calculated
quantity should not  depend on the artificial parameter at all.

Notice that  our approach can be also applied to more general
scheme such as, for instance,  the renormgroup conditions $^{40}$
or the simple equating to zero some corrections:
\begin{eqnarray}{\label{21}}
E_i(\omega_0 ,..., \omega_k)=0, \;\;\;\; i=2,3,...,N.
\end{eqnarray}

\section*{\normalsize 5. Example of application}
As a rule, various schemes of renormalization are illustrated with
describing ground states of the quartic anharmonic oscillator.
Here we consider the less traditional example of obtaining energy
eigenvalues for both ground and excited states of the sextic
anharmonic oscillator with a potential
\begin{eqnarray}{\label{22}}
V(x) = \frac{1}{2} \left(  x^2 + \lambda x^6 \right),
\end{eqnarray}
where we put $\omega=m=1$ for simplicity.

The analytical representation of perturbation expansion for its
energy  obtained by using formulae (\ref{14}) and (\ref{15})
has the form
\begin{eqnarray}{\label{23}}
E_{1}&=&{\frac{1}{2} + n}   ,  \nonumber
\\
E_{3}&=&{2^{-4}} 5 \,   \lambda \left( 1 + 2 n \right) \,
     \left( 3 + 2 n + 2 {n^2} \right)  , \nonumber
\\
E_{5}& =&-  {2^{-8}} \,{{\lambda }^2} \left( 1 + 2 n \right) \,
       \left( 3495 + 4538\,n + 5324\,{n^2} + 1572\,{n^3} +
         786\,{n^4} \right)  , \nonumber
\\
E_{7}&=&2^{-11}5\,{{\lambda }^3}\,\left( 1 + 2\,n \right) \,
     \left( 247935 + 444014\,n + 600050\,{n^2} +
       323868\,{n^3}
\right.
\nonumber
\\
&&\left.
+ 191424\,{n^4} + 35388\,{n^5} +
       11796\,{n^6} \right) .
\end{eqnarray}
This series diverges for any finite value of $\lambda$ and
requires the use of some renormalization procedure. Since  for
discussed potential the above series is, in fact, the expansion in
powers of the Planck constant squared, the trial frequency within
any one-parameter scheme of renormalization   is given by
(\ref{20}) as
\begin{eqnarray}{\label{24}}
\omega^2=\omega_0^2+\omega_2^2 \hbar^2 .
\end{eqnarray}

By taking  use of relations (\ref{19}), the first terms of the renormalized
perturbation
series for the energy eigenvalues then read
\begin{eqnarray}{\label{25}}
E_{1}&=&\omega_0 \left(  {\frac{1}{2}}+n \right)  \nonumber \\
E_{3}&=&2^{-4} \omega_{0}^{-3} \left( 1 + 2\,n \right) \,
     \left[ 5 \lambda \,\left( 3 + 2\,n + 2\,{n^2} \right) \,
        + 4\,{{\omega }_0^2}\,
        \left( 1 - {{\omega }_0^2} \right)  \right]
     \nonumber \\
E_{5}&=&  -2^{-8} \omega_{0}^{-7} \left( 1 + 2\,n \right) \,\times
\nonumber\\
     &&  \times\Big[  {{\lambda }^2} \left( 3495 + 4538\,n + 5324\,{n^2} +
            1572\,{n^3} + 786\,{n^4} \right) \,
           \nonumber\\
& &
+
         120\lambda \,{{\omega }_0^2}\,
          \left( 1 - {{{{\omega }_0}}^2} \right)
         \,\left( 3 + 2\,n + 2\,{n^2} \right) \,
           +
         16\,{{{{\omega }_0}}^4}\,
          {{\left( 1 - {{{{\omega }_0}}^2} \right) }^2}
          \Big]   \nonumber
\\
E_{7}&=&2^{-11}  \omega_{0}^{-11}\,\left( 1 + 2\,n \right)  \times \,
\nonumber\\
  &&
\times \Big[ 5\,{{\lambda }^3} \left( 247935 + 444014\,n + 600050\,{n^2} +
          323868\,{n^3} \right.  \nonumber\\
& & +
\left.
191424\,{n^4} + 35388\,{n^5} +
          11796\,{n^6} \right) \, \nonumber\\
& & +
       28\,        {{\lambda }^2}\,{{{{\omega }_0}}^2}\,
        \left( 1 - {{{{\omega }_0}}^2} \right)  \times
\nonumber \\ && \times
\left( 3495 + 4538\,n + 5324\,{n^2} +
          1572\,{n^3} + 786\,{n^4} \right) \,
\nonumber\\
&  &+
       1200 \lambda \,
        {{{{\omega }_0}}^4}\,
        {{\left( 1 - {{{{\omega }_0}}^2} \right) }^2}
      \,\left( 3 + 2\,n + 2\,{n^2} \right) \, +
       64\,{{{{\omega }_0}}^6}\,
        {{\left( 1 - {{{{\omega }_0}}^2} \right) }^3}
       \Big]
\end{eqnarray}
where $\omega_2$ is excluded due to equation (\ref{24}).

The obtained recursion formulae enable us to calculate
high orders of the renormalized
perturbation series for wide ranges of the coupling
constant $\lambda$ and the radial
quantum number, $n$.

We have computed the partial sums of $N$ corrections  to the
energy eigenvalues  with defining the free parameter $\omega_0$,
order by order under the conditions of the minimal  difference,
$E_N( \omega_0)=0$, and the minimal sensitivity, $( d / d \omega_0
) E_N ( \omega_0 ) = 0 $ and $( d/d \omega_0 ) \sum_{ i=0 }^{ N }
E_i ( \omega_0 ) = 0$.

Contrary to the current view $^{42}$, we recognize that in the
case  of sextic anharmonic oscillator all these schemes appear to
give approximately an equal accuracy.

It has been observed that in the range of values $0 < \lambda \leq
1000$ and \mbox{$0 \leq  n  \leq 1000$} the best accuracy can be
achieved in the limiting case of the weak coupling, for small
$\lambda$, $n$. The obtained energy eigenvalues are most
inaccurate  in the intermediate region   ($\lambda=O(1)$ and $0
\leq n \leq 2$). Further, the Nth-order renormalized approximation
becomes improve as $n$ and $\lambda $ increase.

The typical behaviour of the speed and accuracy of these
one-parameter schemes is illustrated by Table 1. Here the partial
sums of $N$ renormalized corrections to energy eigenvalues derived
under the condition of minimal sensitivity, $(d/d\omega_0)
\sum_{i=0}^{N} E_i (\omega_0) = 0$, are compared with the energy
eigenvalues obtained by the numerical integration of the
Schr{\"{o}}dinger equation. As the numerical integration procedure
it was used the improved shooting method with the Noumerov finite
difference scheme, described in literature $^{43}$, which ensures
the necessary exactness of calculation for the Sturm-Liouville
problems.  Notice that the condition of minimal sensitivity
yields a large number of the trial $\omega_0$-values from which
we choose the one with the flattest \mbox{extremum $^{42, 44}$.}

\section*{\normalsize 6. Conclusion }
A new simple recursion procedure, suitable to applying any
renormalization scheme of improving the perturbation expansions
for the  bound-state problem of
anharmonic oscillator within the framework of the one-dimensional
Schr{\"o}dinger equation, has been proposed.

Based  upon the semiclassical treatment, including $\hbar$-expansions and proper quantization conditions, the
main disadvantage of logarithmic perturbation theory, namely,
the cumbersome description of excited states when nodes of
 wave functions are taking into account, has been avoided.

The derived  formulae have  been adapted
for the use of any renormalization scheme. As a result,
the recursion formulae, obtained for renormalized
perturbation expansions, have the  same simple form
both for  ground and
excited states and provide, in principle, the calculation of
renormalization corrections up to an arbitrary order.

Moreover, the advantage of the method is such flexibility that
within the united approach one can apply various schemes of
renormalization and choose different  values of the trial
parameter depending on the order of perturbation.

The proposed technique has been developed for investigation of the
 bound-state problem within the one-dimensional Schr\"{o}dinger
equation with any potential described by an analytic function
having a simple global minimum. However, taking into account that
the potential having a finite number of local minima becomes, as $\hbar \to 0$, a
sum of infinite deep wells, our procedure may be applyed as well for
investigation of  bound states in a vicinity of a local minimum,
when tunneling can be neglected.

\section*{\normalsize Acknowledgments}
This work was supported in part by the International Soros Science
Education Program (ISSEP) under Grant APU052102.


\begin{table}[ht]\label{table1}
\small
\caption{ The sequences of the partial sums of $N$ renormalized
corrections to the energy eigenvalues of the sextic anharmonic
oscillator with the potential $V(x)=\frac{1}{2} \left(   x^2 +
\lambda x^6 \right)$, computed within the one-parameter scheme of
renormalization under the condition of minimal sensitivity,
$(d/d\omega_0) \sum_{i=0}^{N} E_i (\omega_0) = 0$. The values of
$E_{num}$ were obtained by numerical integration of the
Schr\"{o}dinger equation.
}
\vspace{10pt}
\begin{tabular}{c|ll|ll|ll}
\hline
\vphantom{\Big(}
N & \hss $n=0$ &
& \hss $ n=1$&
&\hss $ n=5$ &\\
\cline{2-7}
\vphantom{\Big(}   & $\lambda=0.01$ & $\lambda=10$ & $\lambda=0.01$
  & $\lambda=10$ & $\lambda=0.01$ & $\lambda=10$ \\
\hline
   1&  0.508693705    &   1.161458  &  1.55611747 & 4.210051 & 6.59434725  &
25.95659\\
   3&  0.508378396    &   1.110292  &  1.55399477 & 4.054344 & 6.57502024  &
25.54466\\
   5&  0.508371342    &   1.104354  &  1.55397174 & 4.047270 & 6.61788050  &
26.44265\\
10&  0.508370692    &   1.102706  &  1.55398991 & 4.056856 & 6.61763448  &
26.42384\\
15&  0.508370674    &   1.102541  &  1.55398998 & 4.057586 & 6.61764001  &
26.42596\\
20&  0.508370673    &   1.102651  &  1.55398999 & 4.057838 & 6.61764261  &
26.42806\\
25&  0.508370689    &  1.102586   &  1.55398995 & 4.057637 & 6.61763908  &
26.42459\\
30&  0.508370674    & 1.102729    &  1.55398996 & 4.057495 & 6.61763913  &
26.42449\\
35&  0.508370726    &   1.102819  &  1.55398997 & 4.057749 & 6.61764005  &
26.42450\\
40&  0.508370676   &   1.102768   &  1.55398996 & 4.057442 & 6.61763907  &
26.42474\\
45&  0.508370682  &  1.102796     &  1.55398996 & 4.057401 & 6.61763907  &
26.42484\\
50&  0.508370681   &  1.102829    &  1.55398996 & 4.057410 & 6.61763906  &
26.42479\\

\hline
\vphantom{\Big(}
$E_{\rm num}$&  0.508370682   &   1.102862& 1.55398996 & 4.057422  &
6.61763908 &
26.42476  \\
\hline
\end{tabular}
\vspace{10pt}
\end{table}

\end{document}